\documentclass[twocolumn,final,times,sort&compress]{elsarticle}
\usepackage{graphicx,amssymb,amsmath,color}
\journal{Physica B}

\begin{document}

\begin{frontmatter}

\title{Magnetization process and low-temperature thermodynamics of a spin-1/2 Heisenberg octahedral chain\tnoteref{grant}}
\tnotetext[grant]{This work was financially supported by the grant of The Ministry of Education, Science, Research and Sport of the Slovak Republic under the contract No. VEGA 1/0043/16 and by the grant of the Slovak Research and Development Agency under the contract No. APVV-14-0073. O.D. and J.R. acknowledge the support by the Deutsche Forschungsgemeinschaft (project RI615/21-2). O.D. was partially supported by Project FF-30F (No. 0116U001539) from the Ministry of Education and Science of Ukraine.}
\author[UPJS]{Jozef Stre\v{c}ka\corref{coraut}} 
\cortext[coraut]{Corresponding author}
\ead{jozef.strecka@upjs.sk}
\author[OVGU]{Johannes Richter}
\author[ICMP,IFNU]{Oleg Derzhko}
\author[ICMP]{Taras Verkholyak}
\author[UPJS]{Katar\'ina Kar\v{l}ov\'a}
\address[UPJS]{Institute of Physics, Faculty of Science, P. J. \v{S}af\'{a}rik University, Park Angelinum 9, 04001 Ko\v{s}ice, Slovakia}
\address[OVGU]{Institut f\"ur Theoretische Physik, Otto-von-Guericke Universit\"at in Magdeburg, 39016 Magdeburg, Germany}
\address[ICMP]{Institute for Condensed Matter Physics, NASU, Svientsitskii Street 1, 79011 L'viv, Ukraine}
\address[IFNU]{Department for Theoretical Physics, Ivan Franko National University of L'viv, Drahomanov Street 12, 79005 L'viv, Ukraine}

\begin{abstract}
Low-temperature magnetization curves and thermodynamics of a spin-1/2 Heisenberg octahedral chain with the intra-plaquette and monomer-plaquette interactions are examined within a two-component lattice-gas model of hard-core monomers, which takes into account all low-lying energy modes in a highly frustrated parameter space involving the monomer-tetramer, localized many-magnon and fully polarized ground states. It is shown that the developed lattice-gas model satisfactorily describes all pronounced features of the low-temperature magnetization process and the magneto-thermodynamics such as abrupt changes of the isothermal magnetization curves, a double-peak structure of the specific heat or a giant magnetocaloric effect.   
\end{abstract}


\end{frontmatter}

\section{Introduction}

One-dimensional quantum Heisenberg spin chains display at low enough temperatures remarkable magnetization curves, which may even basically depend at low magnetic fields on the spin magnitude according to the conjecture made by Haldane \cite{hal83a,hal83b}. Among the most notable features of zero-temperature magnetization curves one could mention fractional magnetization plateaus, quantum spin liquids and macroscopic magnetization jumps, which can be found first of all in frustrated quantum Heisenberg spin models \cite{prl02,hon04}. The macroscopic magnetization jumps emergent at a saturation field are closely connected with flat-band physics \cite{fqh-flat-review,bergholtz,Flach,der15} and they can be alternatively viewed as a condensation of localized magnons \cite{prl02,zhi04,der04,zhi05,der06,der15}. It should be stressed that the localized-magnon picture of the frustrated quantum Heisenberg spin models is of particular importance, because it additionally allows a proper description of low-temperature thermodynamics with the help of simpler classical lattice-gas models projecting out excited states with much higher energy. However, the main drawback of the localized-magnon approach lies in that its validity is usually restricted only to high magnetic fields \cite{zhi04,der04,zhi05,der06,der15}.  

Recently, the localized-magnon approach has been adapted in order to find an exact ground state close to but slightly below saturation field of a spin-$\frac{1}{2}$ Heisenberg octahedral chain, which involves the localized one-magnon state at each elementary square cell (see Fig.~\ref{fig1} for a 
schematic illustration) \cite{str17}. It is worthwhile to remark, moreover, that the spin-$\frac{1}{2}$ Heisenberg octahedral chain exhibits at sufficiently low magnetic fields another exact ground state with the character of the monomer-tetramer phase, which appears due to formation of a singlet state between four spins creating an elementary square plaquette \cite{bos89}. This singlet-tetramer state can be alternatively considered as the localized two-magnon state, which consequently gives us hope for a proper description of low-temperature thermodynamics of the spin-$\frac{1}{2}$ Heisenberg octahedral chain in a full range of the magnetic fields \cite{str17}. In the following, we will develop a novel kind of the localized-magnon approach for the spin-$\frac{1}{2}$ Heisenberg octahedral chain, which indeed provides a consistent description of the low-temperature thermodynamics in a full range of the magnetic fields.     

\section{Spin-$\frac{1}{2}$ Heisenberg octahedral chain}

\begin{figure}
\begin{center}
\includegraphics[width=0.45\textwidth]{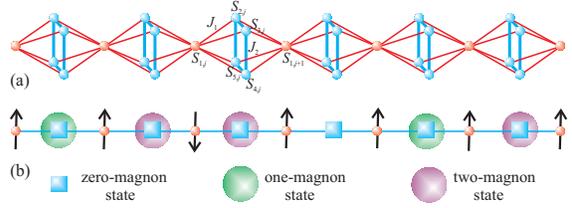}
\end{center}
\vspace{-0.6cm}
\caption{(a) A schematic representation of the spin-$\frac{1}{2}$ Heisenberg octahedral chain. Thick (blue) lines represent the Heisenberg intra-plaquette coupling $J_2$, while thin (red) lines correspond to the monomer-plaquette coupling $J_1$; (b) an equivalent two-component lattice-gas model of hard-core monomers valid in a highly frustrated region $J_2 \geq 2 J_1$. Green and violet balls denote hard-core monomers, which represent one-magnon and two-magnon states of square plaquettes given by Eqs. (\ref{lm1}) and (\ref{lm2}). Unoccupied blue squares denote fully polarized (zero-magnon) state of square plaquettes.}
\label{fig1}
\end{figure}

Let us consider the spin-$\frac{1}{2}$ Heisenberg octahedral chain schematically depicted in Fig.~\ref{fig1}(a) and defined through the Hamiltonian
\begin{eqnarray}
\label{ham}
\hat{\cal H} \!\!&=&\!\! 
\sum_{j=1}^{N} \Bigl[ J_1 (\boldsymbol{\hat{S}}_{1,j} + \boldsymbol{\hat{S}}_{1,j+1}) \!\cdot\! (\boldsymbol{\hat{S}}_{2,j} + \boldsymbol{\hat{S}}_{3,j} + \boldsymbol{\hat{S}}_{4,j} + \boldsymbol{\hat{S}}_{5,j}) \Bigr.  \nonumber \\
\!\!&+&\!\! J_2 (\boldsymbol{\hat{S}}_{2,j}\!\cdot\!\boldsymbol{\hat{S}}_{3,j} + \boldsymbol{\hat{S}}_{3,j}\!\cdot\!\boldsymbol{\hat{S}}_{4,j}
+ \boldsymbol{\hat{S}}_{4,j}\!\cdot\!\boldsymbol{\hat{S}}_{5,j} + \boldsymbol{\hat{S}}_{5,j}\!\cdot\!\boldsymbol{\hat{S}}_{2,j}) \nonumber \\
\Bigl. \!\!&-&\!\! h \sum_{i=1}^{5} \hat{S}_{i,j}^{z} \Bigr],
\end{eqnarray}
where $\boldsymbol{\hat{S}}_{i,j} \equiv (\hat{S}_{i,j}^x, \hat{S}_{i,j}^y, \hat{S}_{i,j}^z)$ stands for a spin-$\frac{1}{2}$ operator at a lattice site given by two subscripts, the former subscript determines a position within the unit cell and the latter subscript the unit cell itself. The parameter $J_1$ denotes the Heisenberg coupling between nearest-neighbour spins from the monomeric and square-plaquette sites, the parameter $J_2$ labels the Heisenberg coupling between nearest-neighbour spins from the same square plaquette and 
the 
Zeeman term $h \geq 0$ refers to a magnetostatic energy of relevant magnetic moments in a magnetic field. The translational invariance is achieved by the choice of a periodic boundary condition $\boldsymbol{S}_{1,N+1} \equiv \boldsymbol{S}_{1,1}$. 

It has been shown in our preceding work that the spin-$\frac{1}{2}$ Heisenberg octahedral chain given by the Hamiltonian (\ref{ham}) can be solved by several complementary analytical and numerical approaches \cite{str17}, whereas a few unconventional quantum ground states can be corroborated even by exact means. In a low-field part $h \leq J_1 + J_2$ of the highly frustrated parameter space $J_2 \geq 2 J_1$ one may for instance employ the variational approach in order to find an exact monomer-tetramer ground state 
\begin{eqnarray}
|{\rm MT} \rangle \!\!=\!\! \prod_{j=1}^N \! |\!\!\uparrow_{1,j}\rangle \!\otimes\! 
\Bigl[\!\frac{1}{\sqrt{3}}(|\!\!\uparrow_{2,j}\downarrow_{3,j}\uparrow_{4,j}\downarrow_{5,j}\rangle \!\!\!\!\!&+&\!\!\!\!\! |\!\!\downarrow_{2,j}\uparrow_{3,j}\downarrow_{4,j}\uparrow_{5,j}\rangle)  \nonumber \\
- \frac{1}{\sqrt{12}} (|\!\!\uparrow_{2,j}\uparrow_{3,j}\downarrow_{4,j}\downarrow_{5,j}\rangle \!\!\!\!\!&+&\!\!\!\!\! |\!\!\uparrow_{2,j}\downarrow_{3,j}\downarrow_{4,j}\uparrow_{5,j}\rangle \nonumber \\
+ |\!\!\downarrow_{2,j}\uparrow_{3,j}\uparrow_{4,j}\downarrow_{5,j}\rangle \!\!\!\!\!&+&\!\!\!\!\! |\!\!\downarrow_{2,j}\downarrow_{3,j}\uparrow_{4,j}\uparrow_{5,j}\rangle) \Bigr]. \nonumber \\
\label{MT}
\end{eqnarray}
On the other hand, the localized-magnon approach \cite{zhi04,der04,zhi05,der06,der15} 
can be adapted in order to afford an exact evidence of the localized many-magnon ground state 
\begin{eqnarray}
|{\rm LM}\rangle = \prod_{j=1}^N \! |\!\!\uparrow_{1,j}\rangle \!\otimes\! \frac{1}{2}
(|\!\!\downarrow_{2,j}\uparrow_{3,j}\uparrow_{4,j}\uparrow_{5,j}\rangle 
\!\!\!\!\!&-&\!\!\!\!\!|\!\!\uparrow_{2,j}\downarrow_{3,j}\uparrow_{4,j}\uparrow_{5,j}\rangle \nonumber \\
+|\!\!\uparrow_{2,j}\uparrow_{3,j}\downarrow_{4,j}\uparrow_{5,j}\rangle
\!\!\!\!\!&-&\!\!\!\!\!|\!\!\uparrow_{2,j}\uparrow_{3,j}\uparrow_{4,j}\downarrow_{5,j}\rangle), \nonumber \\  
\label{LM}
\end{eqnarray}
which appears in the highly frustrated region $J_2 \geq 2 J_1$ at moderate values of the magnetic field $J_1 + J_2 \leq h \leq J_1 + 2J_2$. Of course, 
the classical ferromagnetic state  
\begin{eqnarray}
|{\rm FM} \rangle = \prod_{j=1}^N \! |\!\!\uparrow_{1,j}\uparrow_{2,j}\uparrow_{3,j}\uparrow_{4,j}\uparrow_{5,j}\rangle 
\label{FM}
\end{eqnarray}
becomes an exact ground state for the magnetic fields higher than the saturation value $h_s = J_1+2J_2$. The primary goal of the present work is to develop in the highly-frustrated region $J_2 \geq 2 J_1$ an effective lattice-gas model, which will comprehensively describe the low-temperature magnetization process and thermodynamics.

\section{Lattice-gas model of hard-core monomers}
The many-magnon ground state (\ref{LM}) emergent below the saturation field involves except fully polarized monomeric spins a single localized magnon trapped at each elementary square plaquette as given by the eigenvector  
\begin{eqnarray}
|1\rangle_j = \frac{1}{2} (|\!\!\downarrow_{2,j}\uparrow_{3,j}\uparrow_{4,j}\uparrow_{5,j}\rangle \!\!\!\!\!&-&\!\!\!\!\! |\!\!\uparrow_{2,j}\downarrow_{3,j}\uparrow_{4,j}\uparrow_{5,j}\rangle \nonumber \\
                         + |\!\!\uparrow_{2,j}\uparrow_{3,j}\downarrow_{4,j}\uparrow_{5,j}\rangle \!\!\!\!\!&-&\!\!\!\!\! |\!\!\uparrow_{2,j}\uparrow_{3,j}\uparrow_{4,j}\downarrow_{5,j}\rangle).  
\label{lm1}
\end{eqnarray}
Contrary to this, the monomer-tetramer ground state (\ref{MT}) is constituted by a singlet-tetramer state of the four spins forming an elementary square plaquette, which can be alternatively viewed as the localized two-magnon state given by the eigenvector 
\begin{eqnarray}
|2\rangle_j = 
  \frac{1}{\sqrt{3}}(|\!\!\uparrow_{2,j}\downarrow_{3,j}\uparrow_{4,j}\downarrow_{5,j}\rangle \!\!\!\!\!&+&\!\!\!\!\! |\!\!\downarrow_{2,j}\uparrow_{3,j}\downarrow_{4,j}\uparrow_{5,j}\rangle)  \nonumber \\
- \frac{1}{\sqrt{12}} (|\!\!\uparrow_{2,j}\uparrow_{3,j}\downarrow_{4,j}\downarrow_{5,j}\rangle \!\!\!\!\!&+&\!\!\!\!\! |\!\!\uparrow_{2,j}\downarrow_{3,j}\downarrow_{4,j}\uparrow_{5,j}\rangle \nonumber \\
+ |\!\!\downarrow_{2,j}\uparrow_{3,j}\uparrow_{4,j}\downarrow_{5,j}\rangle \!\!\!\!\!&+&\!\!\!\!\! |\!\!\downarrow_{2,j}\downarrow_{3,j}\uparrow_{4,j}\uparrow_{5,j}\rangle).\!
\label{lm2}
\end{eqnarray}
It is noteworthy that the monomeric spins $S_{1,j}$ are effectively decoupled from the four spins forming the singlet-tetramer state (\ref{lm2}). To summarize, the four spins forming a square plaquette display in the highly frustrated region $J_2 \geq 2J_1$ either the localized one-magnon state (\ref{lm1}) or the localized two-magnon state (\ref{lm2}) or are fully polarized within all available lowest-energy eigenstates of the spin-$\frac{1}{2}$ Heisenberg octahedral chain, while the monomeric spins are always fully polarized except that they are surrounded by two square plaquettes in the singlet-tetramer state (\ref{lm2}) due to the effective decoupling of the monomer-plaquette interaction $J_1$.     

Bearing this in mind, the low-temperature magnetization process and thermodynamics of the spin-$\frac{1}{2}$ Heisenberg octahedral chain can be reformulated as a two-component lattice-gas model of hard-core monomers (see Fig. \ref{fig1}(b) for a schematic illustration), since each square plaquette can host either one localized one-magnon state (\ref{lm1}) represented by the first kind of hard-core monomers with the chemical potential $\mu_1 = J_1 + 2J_2 - h$ or one localized two-magnon state (\ref{lm2}) represented by the second kind of hard-core monomers with the chemical potential $\mu_2 = 2J_1 + 3J_2 - 2h$. The chemical potentials $\mu_1$ and $\mu_2$ of two kinds of the hard-core monomers are determined by an energy difference between the respective localized magnon state (either one- or two-magnon state) and the fully polarized ferromagnetic state serving as a reference state. Consequently, the spin-$\frac{1}{2}$ Heisenberg octahedral chain can be described by the one-dimensional lattice-gas model of hard-core monomers given by the Hamiltonian 
\begin{eqnarray}
{\cal H} = E_{\rm FM}^0 - h \left(2N + \sum_{j=1}^N S_{1,j}^z\right) - \mu_1 \sum_{j=1}^{N} n_{1,j} - \mu_2 \sum_{j=1}^{N} n_{2,j}, \nonumber
\label{elm}
\end{eqnarray} 
where $n_{1,j} = 0,1$ and $n_{2,j} = 0,1$ denote the respective occupation numbers of two kinds of the hard-core monomers with the chemical potentials $\mu_1$ and $\mu_2$, $E_{\rm FM}^0 = N(2J_1 + J_2)$ labels the energy of fully polarized ferromagnetic state in a zero magnetic field and the second term accounts for 
the Zeeman energy. The partition function of the relevant lattice-gas model accounting for all lowest-energy modes of the spin-$\frac{1}{2}$ Heisenberg octahedral chain then follows from the formula
\begin{eqnarray}
{\cal Z} \!\!\!\!\!&=&\!\!\!\!\! {\rm e}^{-\beta E_{\rm FM}^0 + \beta 2Nh} \sum_{\{S_{1,j}^z\}} \prod_{j=1}^N  \sum_{n_{1,j}=0,1} \sum_{n_{2,j}=0,1} 
\!\! (1-n_{1,j}n_{2,j}) \Bigl[ n_{2,j} +   \nonumber \\
(1\!\!\!\!\!&-&\!\!\!\!\!n_{2,j})(\textstyle\frac{1}{2} + S_{1,j}^z)(\textstyle\frac{1}{2} + S_{1,j+1}^z) \Bigr] {\rm e}^{\beta [\mu_1 n_{1,j} + \mu_2 n_{2,j} + \frac{h}{2} (S_{1,j}^z + S_{1,j+1}^z)]}, \nonumber \\
\label{z}
\end{eqnarray}
where $\beta = 1/(k_{\rm B} T)$, $k_{\rm B}$ is Boltzmann's constant, $T$ is the absolute temperature and the prefactor $(1-n_{1,j}n_{2,j})$ establishes a hard-core constraint for both kinds of the monomeric particles. The expression in square brackets 
ensures the paramagnetic character of the monomeric spins once the Heisenberg spins from two neighboring square plaquettes are in the singlet-tetramer state (\ref{lm2}) and, respectively, the fully polarized nature of the monomeric spins if the Heisenberg spins from two neighboring square plaquettes are either fully polarized or fall into the localized one-magnon state (\ref{lm1}). After the summation over the occupation numbers $n_{1,j}$ and $n_{2,j}$ of the hard-core monomers is explicitly carried out, the expression emerging in Eq. (\ref{z}) after the product symbol depends only on spin states of two nearest-neighbour monomeric spins and hence, it can be further interpreted as the transfer matrix
\begin{eqnarray}
T_{S_{\!1,j}^z, S_{\!1,j+1}^z}\!\! = {\rm e}^{\frac{\beta h}{2} (S_{\!1,j}^z + S_{\!1,j+1}^z)} 
\Bigl[(\textstyle\frac{1}{2} \!+\! S_{\!1,j}^z)(\textstyle\frac{1}{2} \!+\! S_{\!1,j+1}^z)(1 \!+\! {\rm e}^{\beta \mu_1}) \!+\! {\rm e}^{\beta \mu_2}\Bigr].
\nonumber
\end{eqnarray}
The partition function then readily follows from the standard transfer-matrix approach \cite{bax82}
\begin{eqnarray}
{\cal Z} \!\!\!\!\!&=&\!\!\!\!\!  {\rm e}^{-\beta E_{\rm FM}^0 + \beta 2Nh} \sum_{\{S_{1,j}^z\}} \prod_{j=1}^N  T_{S_{1,j}^z, S_{1,j+1}^z} \nonumber \\
         \!\!\!\!\!&=&\!\!\!\!\!  {\rm e}^{-\beta E_{\rm FM}^0 + \beta 2Nh} \, \mbox{Tr} \, T^N  = {\rm e}^{-\beta E_{\rm FM}^0 + \beta 2Nh} (\lambda_{+}^N + \lambda_{-}^N),
\label{ztm}
\end{eqnarray}
where $\lambda_{\pm}$ denote two transfer-matrix eigenvalues
\begin{eqnarray}
\lambda_{\pm} \!\!\!\!\!&=&\!\!\!\!\! \frac{1}{2} \Biggl\{ {\rm e}^{\frac{\beta h}{2}} (1 + {\rm e}^{\beta \mu_1} + {\rm e}^{\beta \mu_2}) + {\rm e}^{-\frac{\beta h}{2} + \beta \mu_2} \nonumber \\
\!\!\!\!\!&\pm&\!\!\!\!\! \sqrt{[{\rm e}^{\frac{\beta h}{2}} (1 + {\rm e}^{\beta \mu_1} + {\rm e}^{\beta \mu_2}) - {\rm e}^{-\frac{\beta h}{2} + \beta \mu_2}]^2 + 4 {\rm e}^{2 \beta \mu_2}} \Biggr\}.
\label{tme}
\end{eqnarray}
The Helmholtz free energy per spin can be calculated in the thermodynamic limit $N \to \infty$ from the larger transfer-matrix eigenvalue 
\begin{eqnarray}
f \!\!\!\!\!&=&\!\!\!\!\! - k_{\rm B} T \lim_{N \to \infty} \frac{1}{5N} \ln {\cal Z} \nonumber \\
  \!\!\!\!\!&=&\!\!\!\!\! \frac{1}{5} (2J_1 + J_2) - \frac{2h}{5}  - \frac{1}{5} k_{\rm B} T \ln \lambda_{+},
\label{lmgfe}
\end{eqnarray}
whereas the magnetization, entropy and specific heat can be then computed from Eq. (\ref{lmgfe}) by making use of basic thermodynamic relations
\begin{eqnarray}
m = -\frac{\partial f}{\partial h}, \qquad s = -\frac{\partial f}{\partial T}, \qquad  c = - T \frac{\partial^2 f}{\partial T^2}.
\label{msc}
\end{eqnarray}

\begin{figure*}[thb]
\begin{center}
\includegraphics[width=0.46\textwidth]{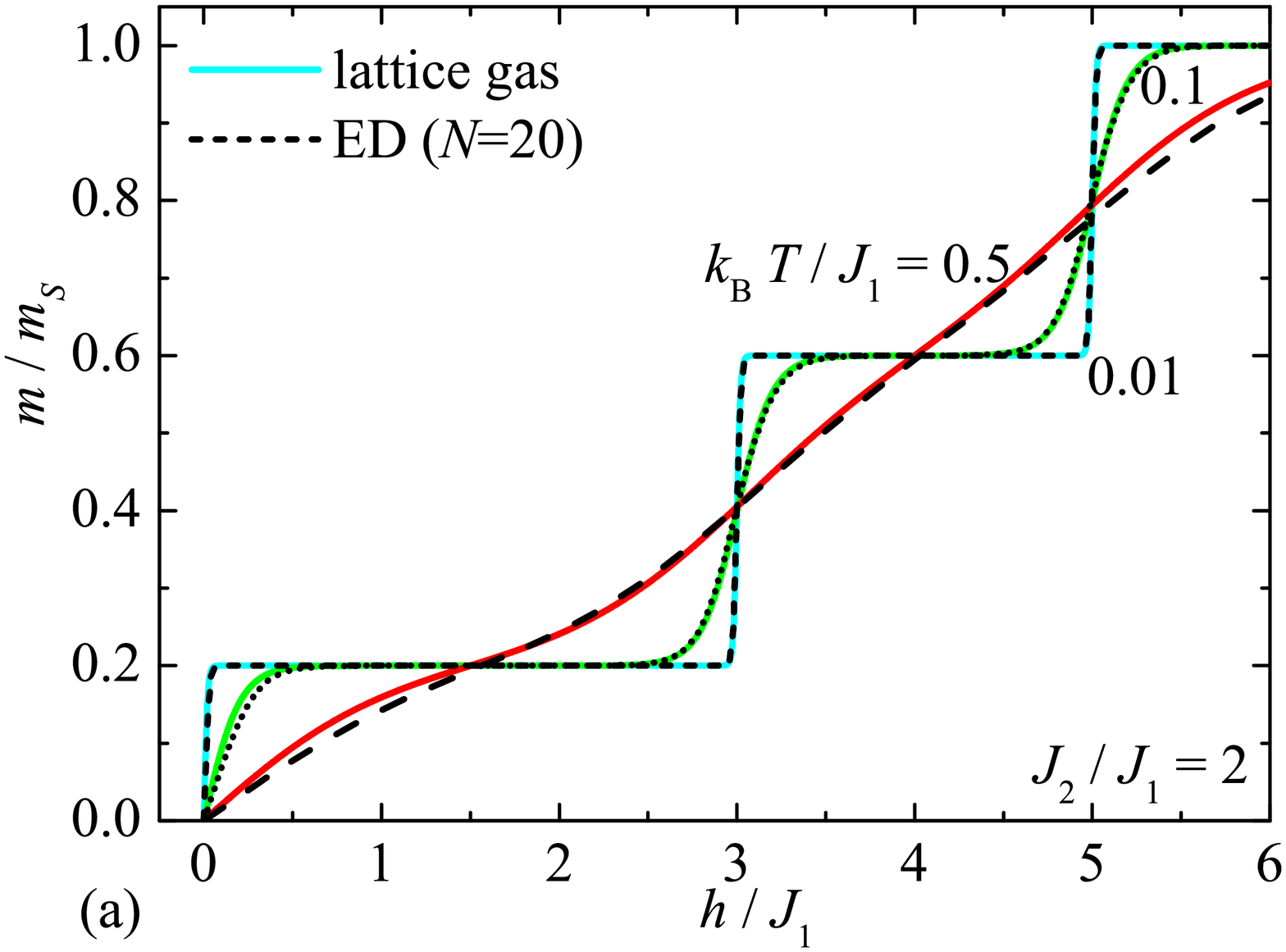}
\hspace{0.2cm}
\includegraphics[width=0.48\textwidth]{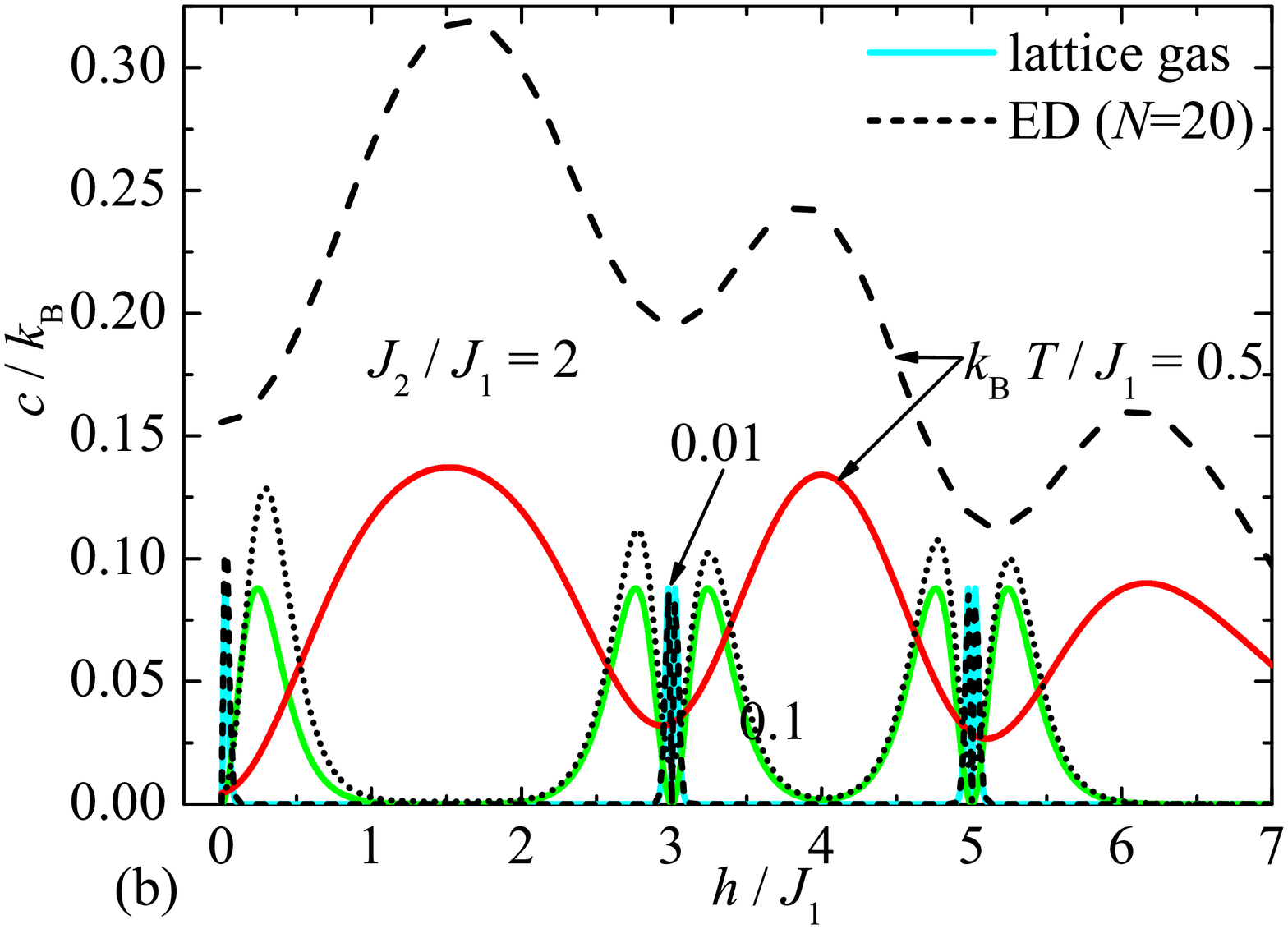}
\includegraphics[width=0.46\textwidth]{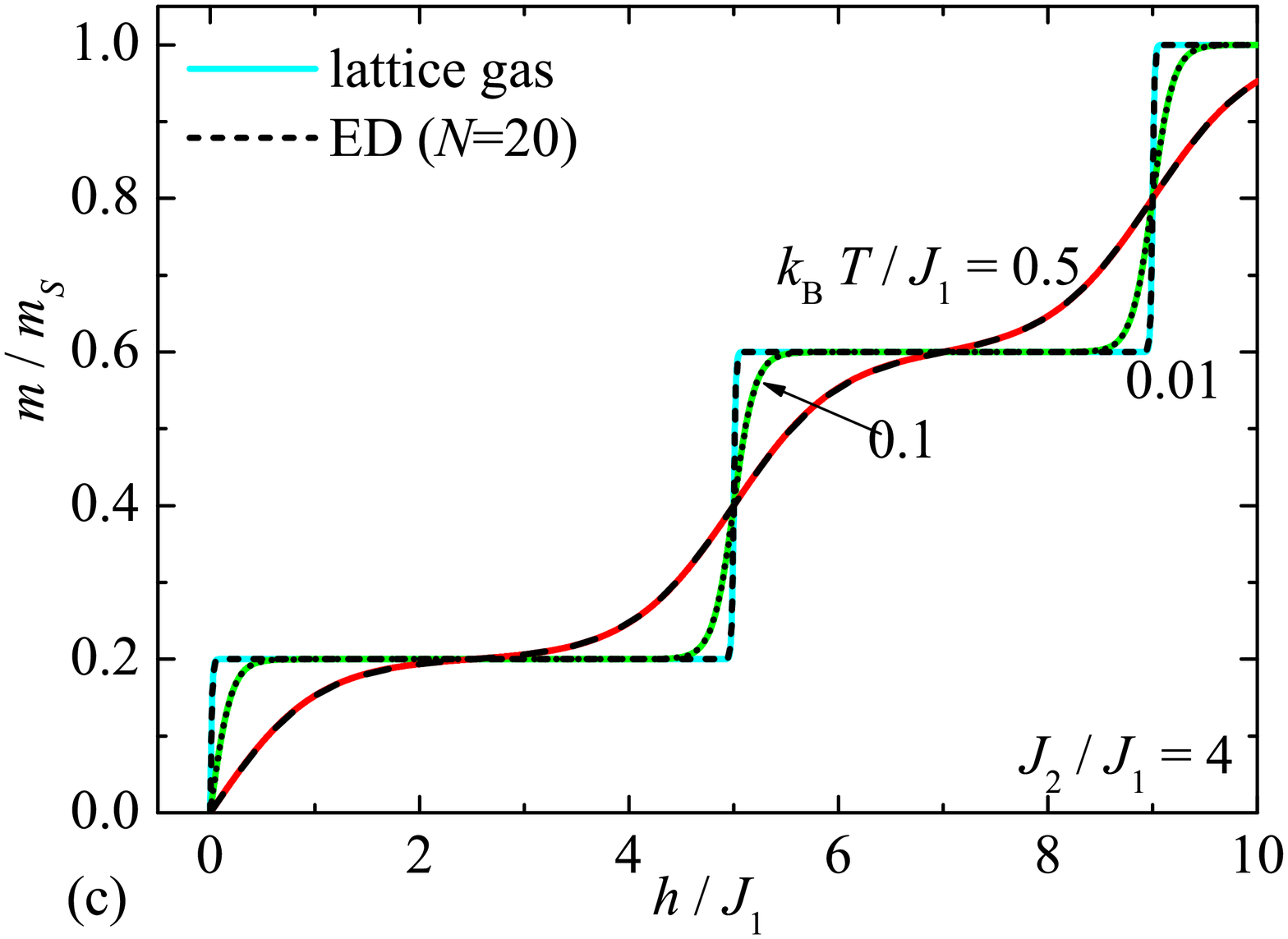}
\hspace{0.2cm}
\includegraphics[width=0.48\textwidth]{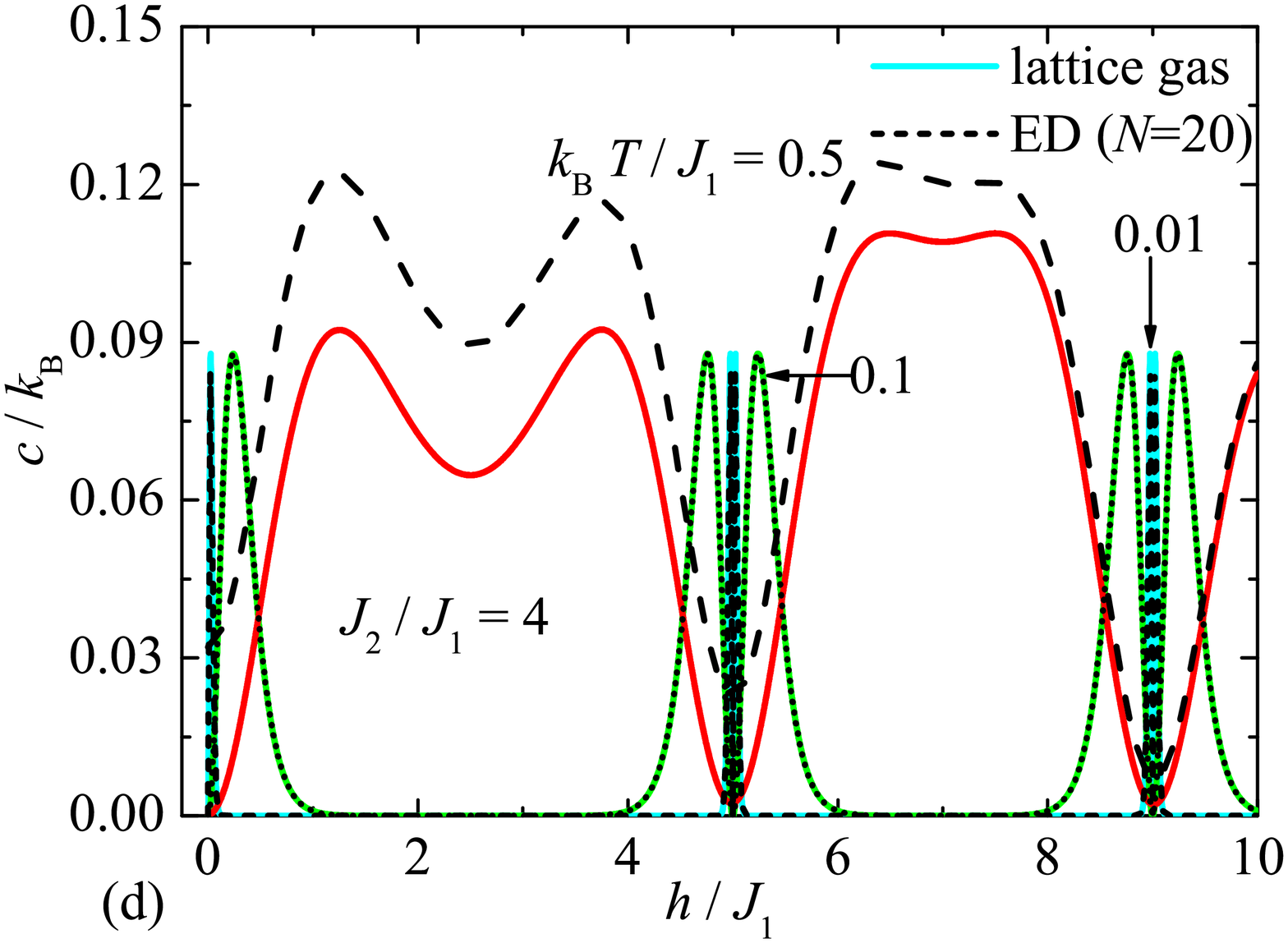}
\end{center}
\vspace{-0.8cm}
\caption{The magnetization and the specific heat of the spin-$\frac{1}{2}$ Heisenberg octahedral chain as a function of the magnetic field for two different values of the interaction ratio: (a)-(b) $J_2/J_1 = 2$; (c)-(d) $J_2/J_1 = 4$. Solid lines display the magnetization and specific-heat data derived from the free energy (\ref{lmgfe}) of the effective lattice-gas model in the thermodynamic limit $ N \to \infty$, while broken lines of different styles illustrate the full ED data for the spin-$\frac{1}{2}$ Heisenberg octahedral chain of $L = 20$ spins ($N=4$ unit cells).}
\label{fig2}
\end{figure*}

\section{Results and discussion}

The magnetization and specific-heat data acquired from the free energy (\ref{lmgfe}) of the effective lattice-gas 
model according to Eqs. (\ref{msc}) are plotted in Fig. \ref{fig2} against the magnetic field 
for three different values of temperature together with 
exact diagonalization (ED)  data of the spin-$\frac{1}{2}$ Heisenberg 
octahedral chain, which serve as a useful benchmark for the developed localized-magnon approach.
To perform the ED calculations we used J. Schulenburg's {\it spinpack} code
\cite{spinpack}. Fig. \ref{fig2}(a)-(b) shows the isothermal dependence of the magnetization and specific heat on the magnetic field for an extreme case $J_2/J_1 = 2$ delimiting the applicability of the localized-magnon approach. It is therefore quite surprising that ED data for the magnetization are in a perfect agreement with the developed localized-magnon approach up to the moderate temperature $k_{\rm B} T/J_1 \approx 0.5$, while the specific heat data start to deviate from ED data at much lower temperatures $k_{\rm B} T/J_1 \approx 0.1$. On the other hand, it can be seen from Fig. \ref{fig2}(c)-(d) that both magnetization as well as specific-heat data are in a perfect coincidence with ED data when the interaction ratio $J_2/J_1=4$ falls deeper into 
the range of the applicability of the localized-magnon approach. It could be thus concluded that the effective lattice-gas model correctly reproduces a stepwise magnetization curve with intermediate plateaus at one-fifth and three-fifths of the saturation magnetization related to the monomer-tetramer (\ref{MT}) and localized many-magnon (\ref{LM}) ground states, as well as, the marked field dependence of the specific heat with two local maxima emergent around each critical field associated with the magnetization jump. It is noteworthy that the developed localized-magnon approach reproduces the ED data the more reliably, the greater the interaction ratio $J_2/J_1$ is [cf. Fig.~\ref{fig2}(a)-(b) with Fig.~\ref{fig2}(c)-(d)].

\begin{figure*}[thb]
\begin{center}
\includegraphics[width=0.51\textwidth]{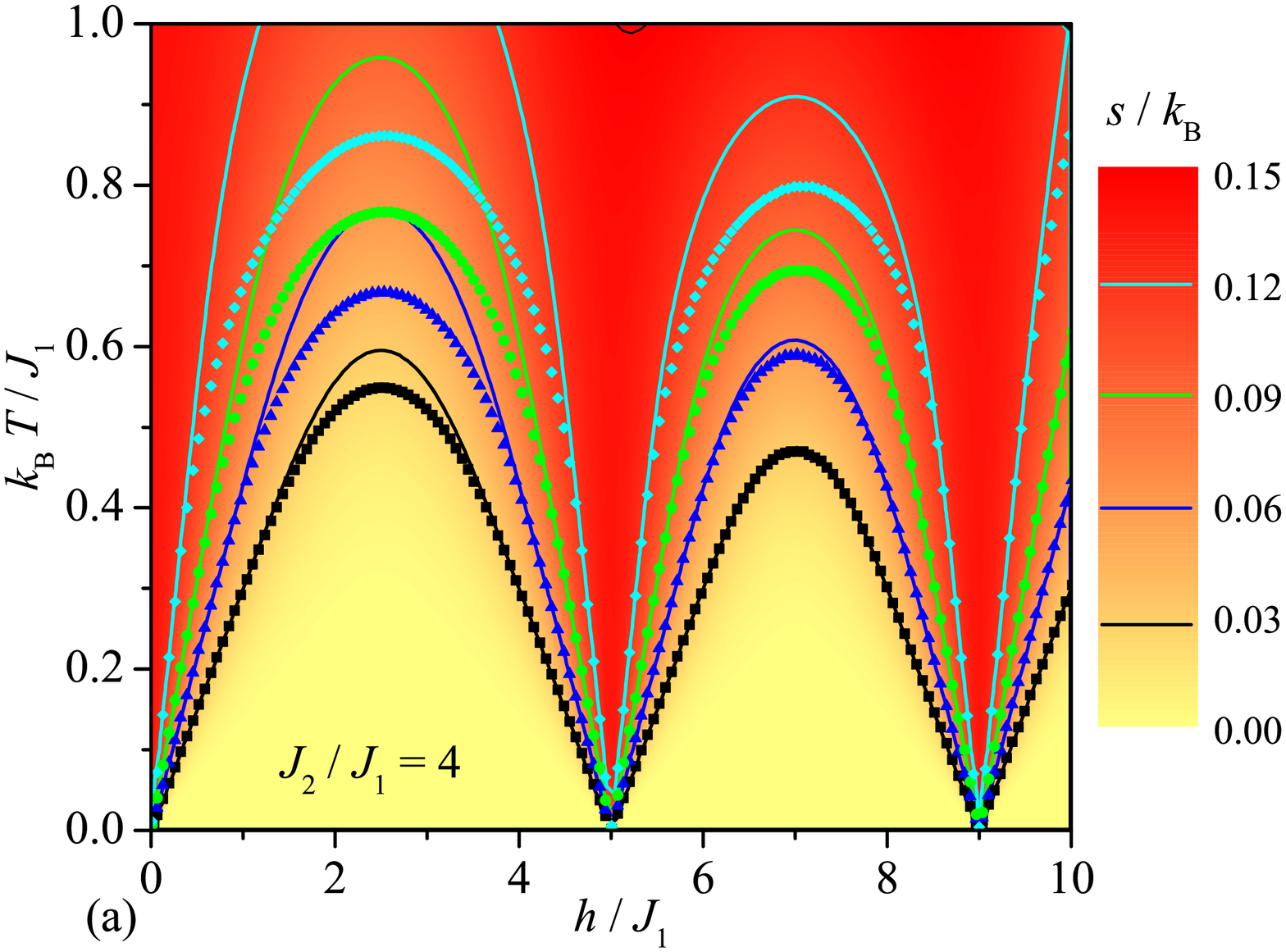}
\hspace{-0.5cm}
\includegraphics[width=0.51\textwidth]{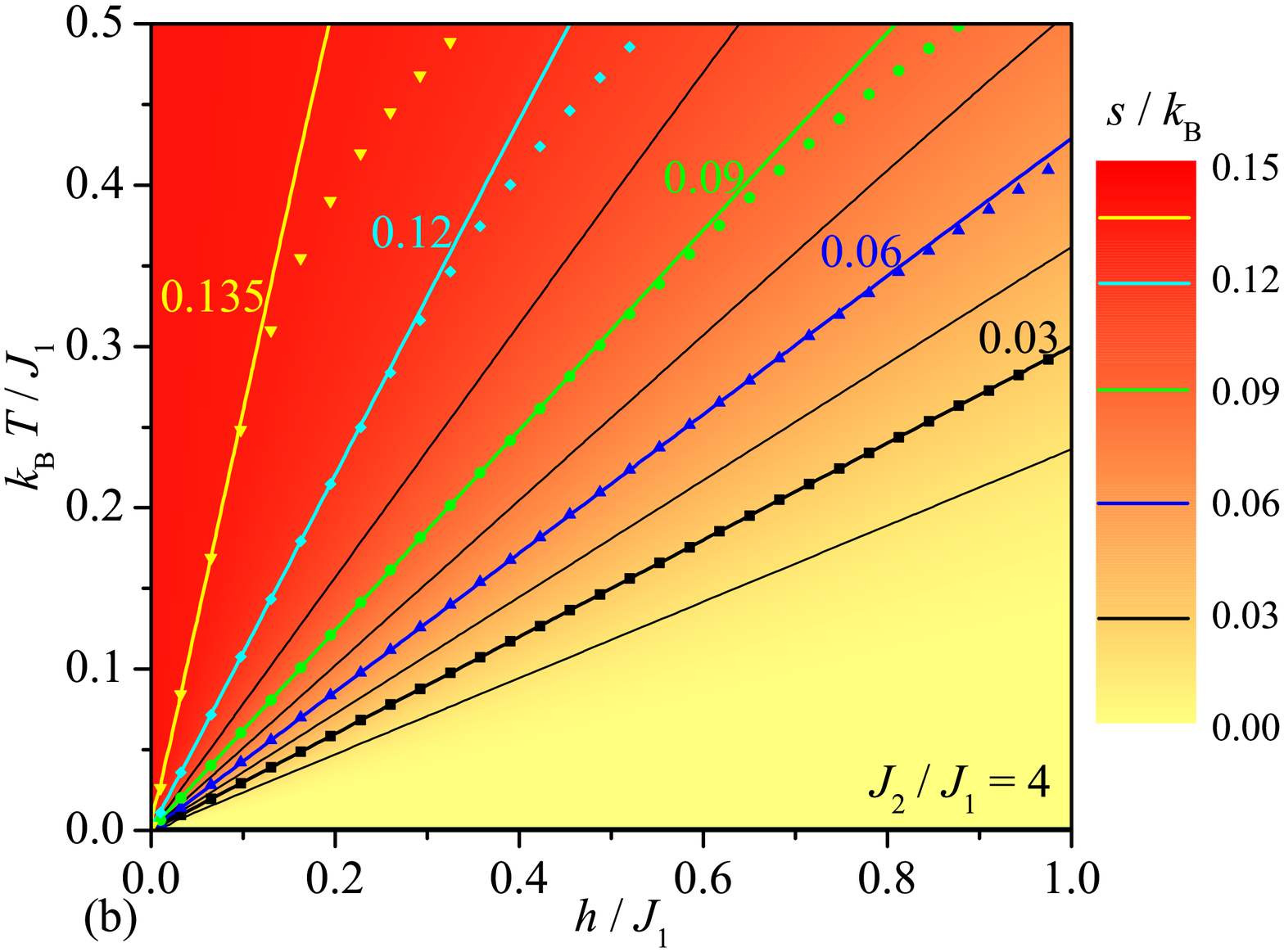}
\end{center}
\vspace{-0.8cm}
\caption{A density plot of the entropy per spin in the magnetic field-temperature plane for the interaction ratio $J_2/J_1 = 4$. Isoentropy (solid) lines calculated from the free energy (\ref{lmgfe}) of the effective lattice-gas model in the thermodynamic limit $ N \to \infty$, which demonstrate adiabatic change of temperature achieved upon variation of the magnetic field, are confronted with ED data (symbols) for the spin-$\frac{1}{2}$ Heisenberg octahedral chain of $L = 20$ spins ($N=4$ unit cells).}
\label{fig3}
\end{figure*}

Next, the density plot of the entropy per spin as obtained from the effective lattice-gas model is displayed in Fig. \ref{fig3} for
a particular value  of the interaction ratio $J_2/J_1 = 4$. The isoentropy lines, which are shown in Fig. \ref{fig3} as solid lines, bring insight into the temperature response of the spin-$\frac{1}{2}$ Heisenberg octahedral chain achieved upon variation of the magnetic field under 
adiabatic conditions. For comparison, Fig. \ref{fig3} also contains the isoentropy data of the spin-$\frac{1}{2}$ Heisenberg octahedral chain as obtained from the ED method, see symbols of different styles and colors. It is quite apparent that one may detect in a low-field region a steep decline of temperature upon diminishing of the magnetic field whenever the entropy per spin is fixed sufficiently close to its residual value $s = k_{\rm B} \ln(2) / 5 \approx 0.1386 k_{\rm B}$ closely connected with a paramagnetic character of the monomeric spins at zero magnetic field. From this perspective, experimental realizations of the spin-$\frac{1}{2}$ Heisenberg octahedral chain falling into the highly-frustrated region $J_2/J_1 \geq 2$ could provide sought after materials for an efficient low-temperature refrigeration, since temperature drops down rapidly near zero magnetic field due to a giant 
magnetocaloric effect, see also \cite{cool1,cool2,cool3}. 

\section{Conclusion}

In the present work we have introduced a classical two-component lattice-gas model of hard-core monomers, which satisfactorily describes the low-temperature magnetization process and thermodynamics of the spin-$\frac{1}{2}$ Heisenberg octahedral chain in the highly frustrated parameter space $J_2/J_1 \geq 2$. The developed classical lattice-gas model was built from localized one-magnon and two-magnon eigenstates (\ref{lm1}) and (\ref{lm2}) of elementary square plaquettes, which are essential for the monomer-tetramer (\ref{MT}) and localized many-magnon (\ref{LM}) ground states emergent in the highly frustrated region $J_2/J_1 \geq 2$, as well as, their low-lying excited states. It has been demonstrated that the classical description of magneto-thermodynamic quantities (e.g. magnetization, entropy and specific heat) based on the two-component lattice-gas model of hard-core monomers is reliable up to moderate temperatures and it does not fail neither at field-driven quantum phase transitions accompanied with the magnetization jump. Our future goal is to extend a validity of the present classical description of magneto-thermodynamics beyond the highly frustrated parameter region $J_2/J_1 < 2$.

\end{document}